\RequirePackage{lineno}
\documentclass[a4paper]{jpconf}
\usepackage{graphicx}
\begin{document}
\title{New Results from ArgoNeuT}

\author{Tingjun Yang for the ArgoNeuT Collaboration}

\address{Fermilab, Batavia, IL, USA}

\ead{tjyang@fnal.gov}

\begin{abstract}
In this article, I discuss the latest results from ArgoNeuT, including inclusive charged current cross sections and studies of final-state interactions.
\end{abstract}

\section{Introduction}
Liquid argon time projection chambers (LArTPCs) offer fine-grained tracking and precise calorimetry, which are ideal for the study of neutrino interactions and search for rare phenomena such as proton decay. The ArgoNeuT (Argon Neutrino Teststand) experiment at Fermilab \cite{Anderson:2012vc} was the first experiment that utilizes a LArTPC exposed to a low-energy neutrino beam (neutrino energies in the 0.5-10.0 GeV range). In this article, I will discuss the latest results from ArgoNeuT. 

\section{ArgoNeuT}
ArgoNeuT consists of a vacuum insulated cryostat for ultra-pure liquid argon containment, in which is mounted a time projection chamber (TPC). The cryostat volume is 550 L (0.77 t). The TPC is 47 cm wide (drift direction), 40 cm high and 90 cm long (neutrino beam direction), corresponding to an active volume of 170 L. A high voltage is applied to the cathode to produce a uniform electric field of 500 V/cm inside the TPC. The electron drift velocity is 1.59 mm/$\mu$s at this nominal field, which leads to a maximum drift time of 295 $\mu$s. The ionization electrons are read out by two wire planes (induction and collection planes). 

ArgoNeuT took data in the NuMI (Neutrinos at the Main Injector) beam from September 2009 to February 2010. ArgoNeuT was located approximately 1.5 m upstream of the MINOS Near Detector (MINOS-ND), with the TPC centered 26 cm below the center of the NuMI on-axis beam. The MINOS-ND was used as a spectrometer to measure the momentum and charge of uncontained long-track muons from charged current (CC) neutrino interactions in the ArgoNeuT. The physics run consisted of about two weeks of neutrino-mode running and four-and-a-half months of anti-neutrino mode running. 

\section{Inclusive CC cross sections}
Inclusive neutrino cross sections have been measured on a variety of targets, however, measurements on the liquid argon target are sparse. Ref.~\cite{Anderson:2011ce} presents the first measurement of inclusive muon neutrino CC differential cross sections on argon using the ArgoNeuT data taken in the {\it neutrino-mode} beam. We are currently finalizing similar measurements of inclusive muon neutrino and anti-neutrino CC differential cross sections on argon using the ArgoNeuT data taken in the {\it anti-neutrino-mode} beam. This corresponds to data from the $1.20\times10^{20}$ protons on target (POT), for which MINOS and ArgoNeuT were both up and running over the 4.5 month physics run. Neutrinos in the anti-neutrino beam originate from the decay of unfocused $\pi^{+}$s and have a broad energy spectrum. Their interactions comprise almost half of all neutrino/anti-neutrino-induced events in the detector, which enables the measurements of both $\nu_{\mu}$ and $\bar{\nu}_{\mu}$ CC differential cross sections using this data sample. 

The differential cross sections are measured in terms of muon angle and muon momentum. Neutrino events are reconstructed in the framework of the LArSoft automated reconstruction software package. The muon angle is measured in the TPC. The muon momentum and the sign of the muon charge are measured in the magnetized MINOS-ND. After three dimensional track formation in ArgoNeuT, an attempt is made to match the tracks that leave the ArgoNeuT TPC with muons that have been reconstructed in MINOS and have a hit within 20 cm of upstream face of the detector. 

\begin{figure}
\begin{center}
\includegraphics[width=30pc]{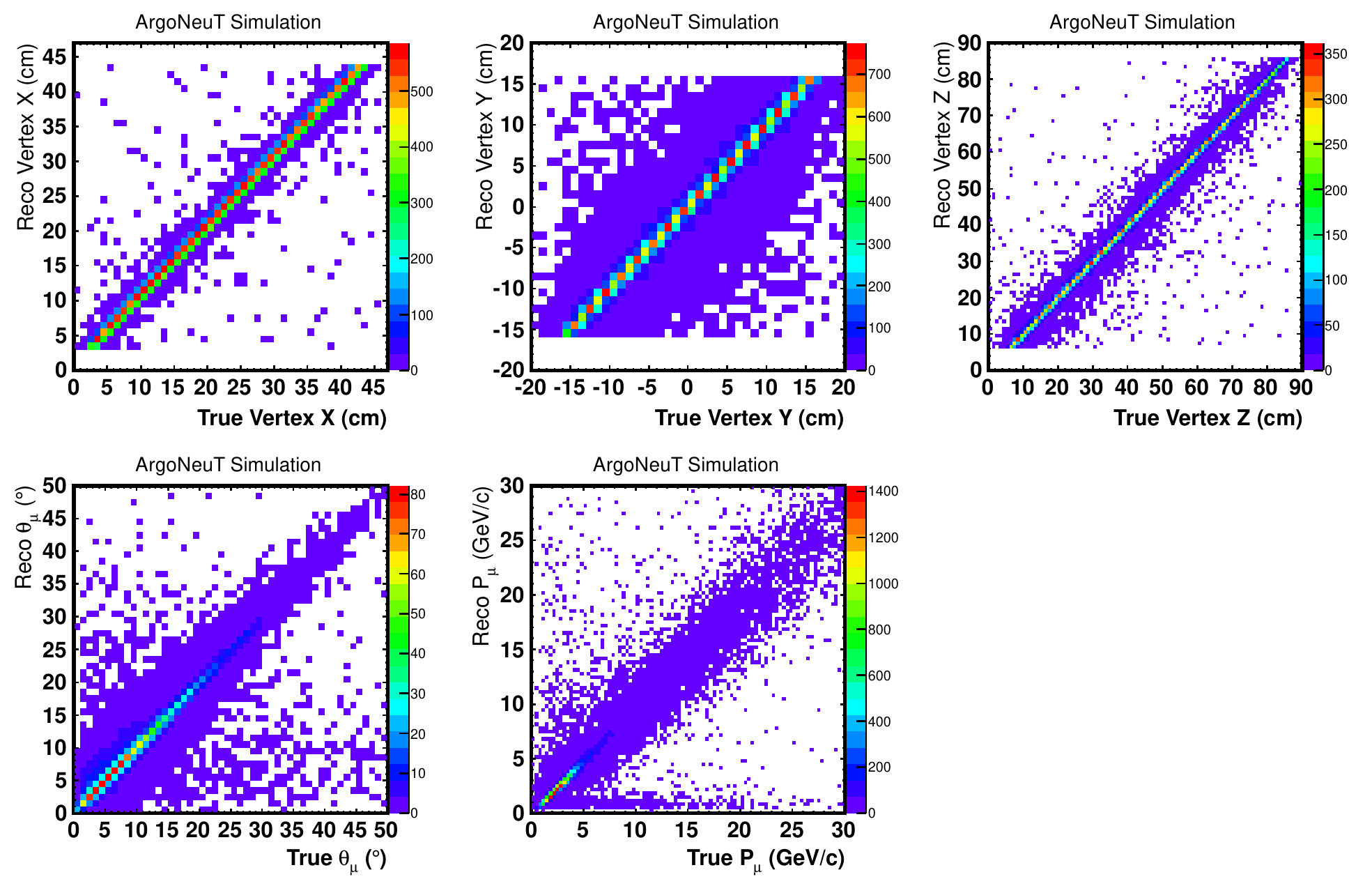}
\end{center}
\caption{\label{recotruth}Comparisons of reconstructed quantities (vertex x, y, z, muon angle and muon momentum) versus true quantities for simulated $\nu_{\mu}$s and $\bar{\nu}_{\mu}$s after event selection.}
\end{figure}

The neutrino event's reconstructed vertex is required to be inside of the ArgoNeuT fiducial volume. The track matching criteria along with a requirement that the reconstructed and matched MINOS track is negatively (positively) charged represent the only other selection criteria for CC $\nu_{\mu}$s ($\bar{\nu}_{\mu}$s) in this analysis. The neutral current (NC) background is highly suppressed by the track matching requirement since the probability of reconstructing tracks from hadrons entering MINOS is low. The remaining NC background and wrong-sign (WS) background are estimated using a simulated sample. The WS background is defined as negative (positive) muons from the CC interactions matched to positive (negative) muons in MINOS because of the misreconstruction of the muon sign in MINOS. Neutrino-induced crossing muons that pass the fiducial volume requirement due to inefficiency in vertex reconstruction are removed through handscanning of events. Figure~\ref{recotruth} shows the comparisons of reconstructed quantities versus true quantities for simulated $\nu_{\mu}$s and $\bar{\nu}_{\mu}$s after event selection. The event reconstruction is unbiased and reliable even for complicated events such as deep inelastic scatterings (DIS). 

\begin{figure}
\begin{center}
\includegraphics[width=15pc]{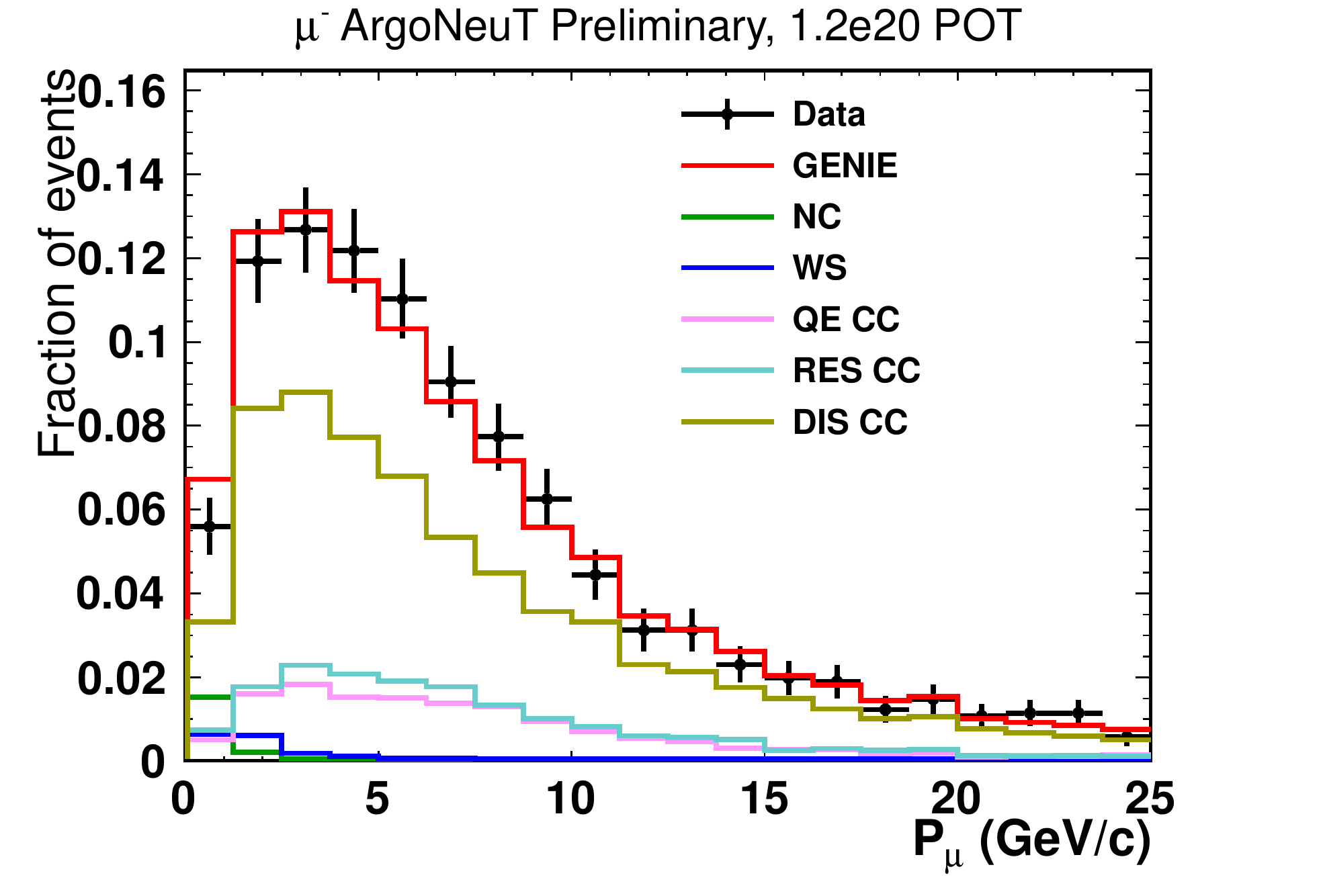}
\includegraphics[width=15pc]{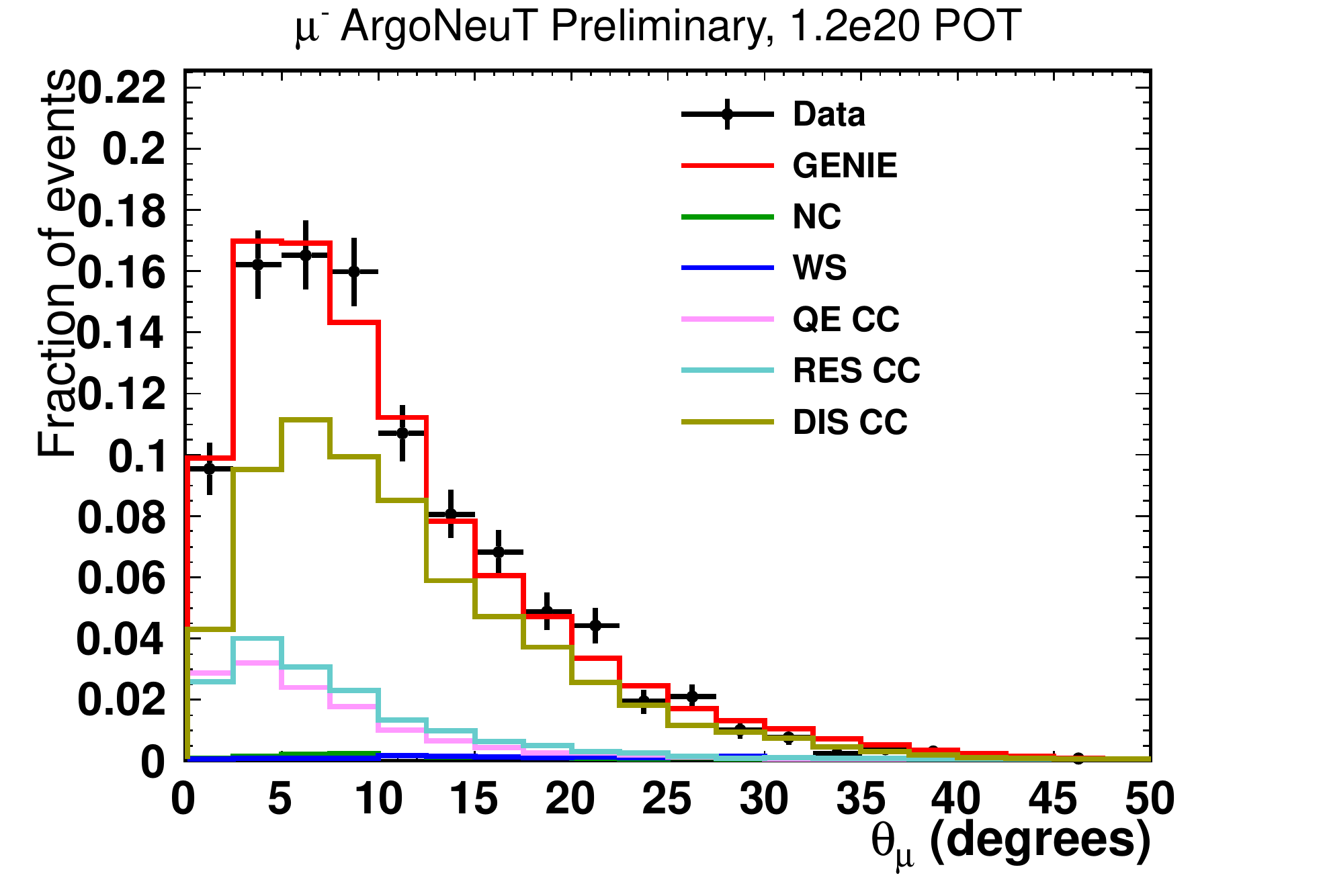}
\includegraphics[width=15pc]{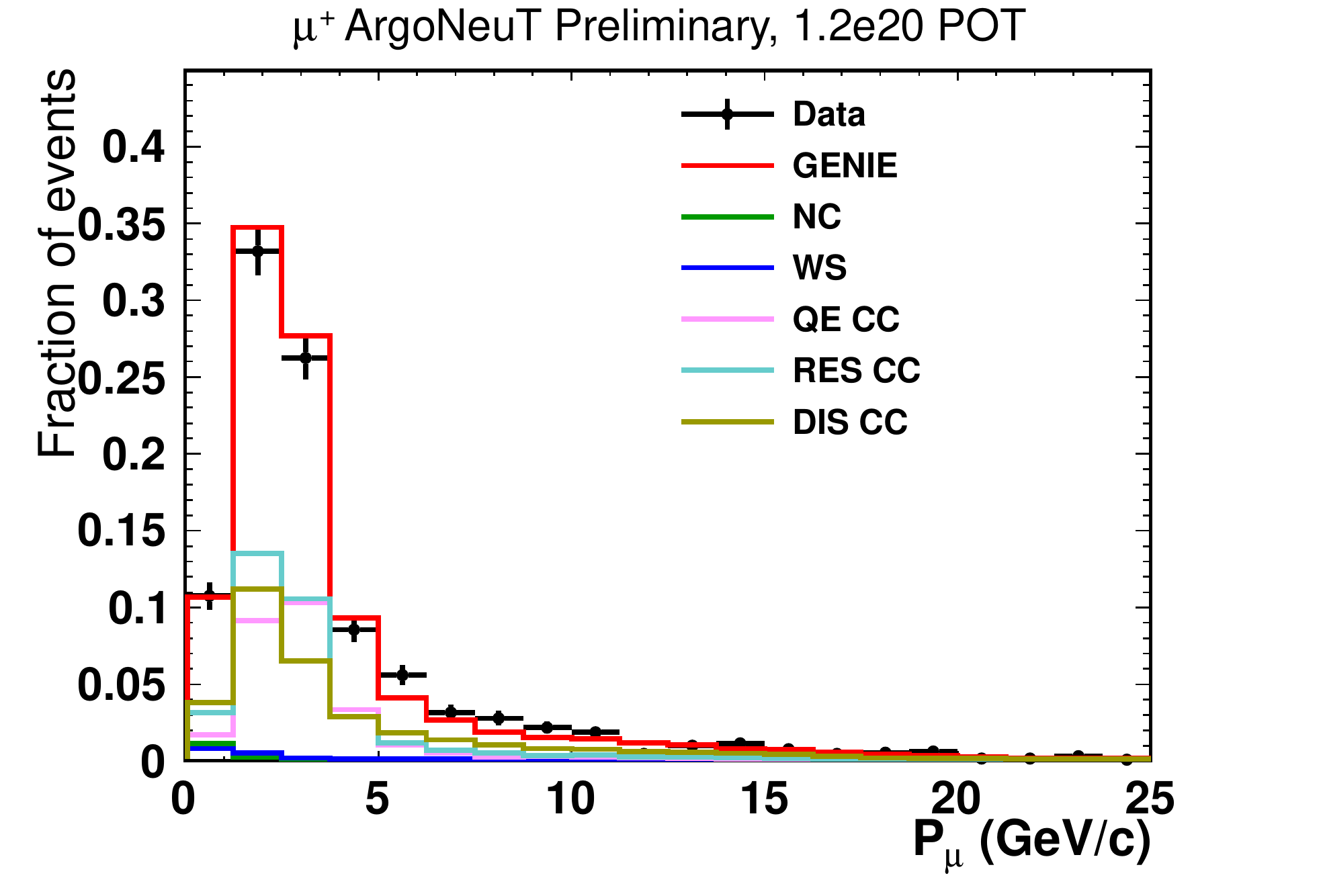}
\includegraphics[width=15pc]{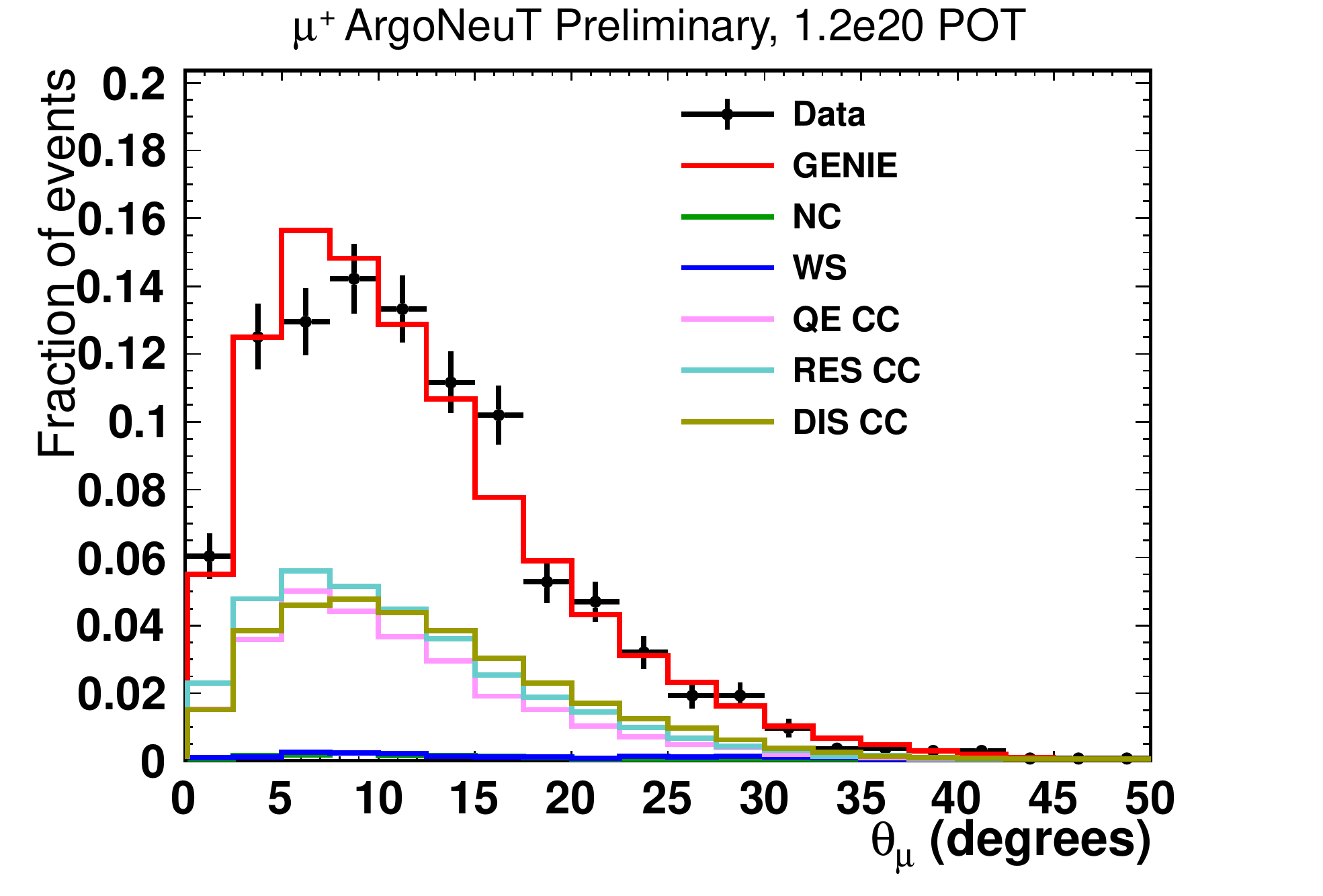}
\end{center}
\caption{\label{ccinc}Distributions of muon angle and muon momentum for $\mu^{-}$'s and $\mu^{+}$'s compared with {\sc genie} predictions. The {\sc genie} prediction is broken down as the NC (neutral current) background, WS (wrong-sign) background, QE (quasielastic) CC, RES (resonance) CC and DIS (deep inelastic scattering) CC. Both data and {\sc genie} total predictions are normalized to unit area.}
\end{figure}

Figure~\ref{ccinc} shows distributions of  muon angle and muon momentum for $\mu^{-}$'s and $\mu^{+}$'s after full event selection compared with {\sc genie} \cite{Andreopoulos:2009rq} predictions. Both data and {\sc genie} total predictions are normalized to unit area. The {\sc genie} simulation describes the shape of data distributions well. The NC and WS backgrounds are negligible. The $P_{\mu^{-}}$ distribution is broader than the $P_{\mu^{+}}$ distribution because of the broader $\nu_{\mu}$ energy spectrum. The $\nu_{\mu}$ sample is dominated by DIS interactions while the $\bar{\nu}_{\mu}$ sample has approximately equal contributions from quasielastic scattering (QE), resonance production (RES) and DIS. The final differential cross sections after correcting for detector acceptance and smearing, selection efficiency, and accounting for neutrino flux will be released soon. 

\section{Topological cross sections}
Conventionally measurements of exclusive channels are performed by looking for a certain event topology in the detector. For example, an event with a muon and a proton is a candidate for $\nu_{\mu}$ CCQE interaction. However, nuclear effects play an important role in neutrino-nucleus interactions. Intranuclear rescattering, a.k.a final state interaction (FSI), and possible effects of correlation between target nucleons can change event topology drastically. For example, the proton produced in a $\nu_{\mu}$ CCQE interaction can be absorbed before it exits the argon nucleus so the event only has a muon, which does not look consistent with a CCQE interaction. On the other hand, a pion produced in a resonance production can be absorbed so the event has a muon and a proton, which looks consistent with a CCQE interaction. Conventional exclusive-channel measurements rely on neutrino generators (e.g. {\sc genie}) to correct for nuclear effects, which makes the results model dependent. In ArgoNeuT we developed a different approach: instead of measuring a specific channel, we measure the cross section for a well-defined event topology \cite{Palamara:2013maa}. Such measurements provide useful information for both neutrino cross sections and nuclear effects. 

A first topological analysis is currently developed by the ArgoNeuT experiment, namely $1\mu+Np+0\pi$, where we select events with one muon, any number of protons but without pions. The analysis takes two steps. First a similar event selection to the inclusive CC cross section measurements is applied to select events with muons. The selected events are then scanned to remove pion background and reconstruct the proton kinematics. The particle identification is done taking advantage of the detailed calorimetry information provided by the LArTPC. The particle species is determined by the energy deposition information as a function of residual range along the track trajectory if the particle stops inside the TPC. The kinetic energy is measured by summing the charge collected on each wire after correcting for electron lifetime and recombination effects. The threshold to reconstruct a proton in ArgoNeuT is $T_{p} = 21$ MeV, where $T_{p}$ is the proton kinetic energy.

\begin{figure}
\begin{center}
\includegraphics[width=15pc]{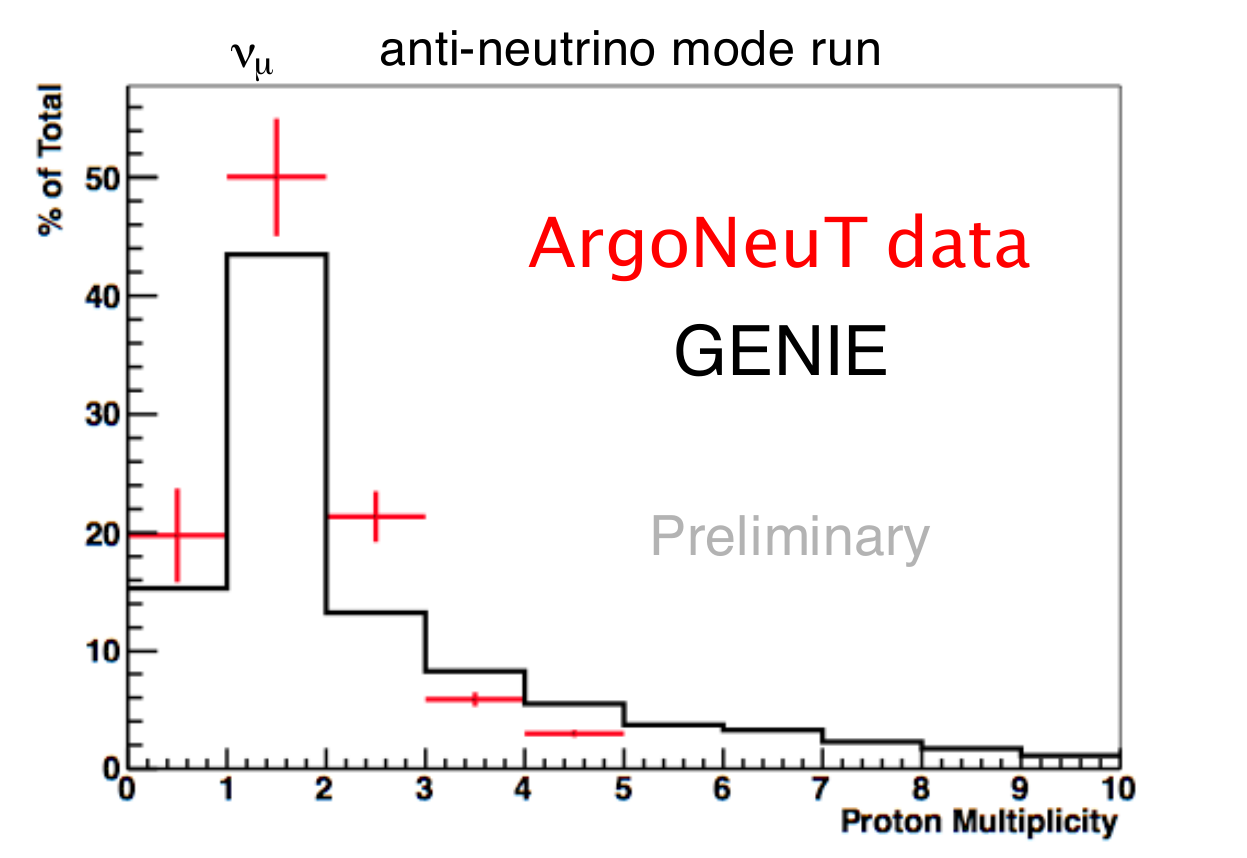}
\includegraphics[width=15pc]{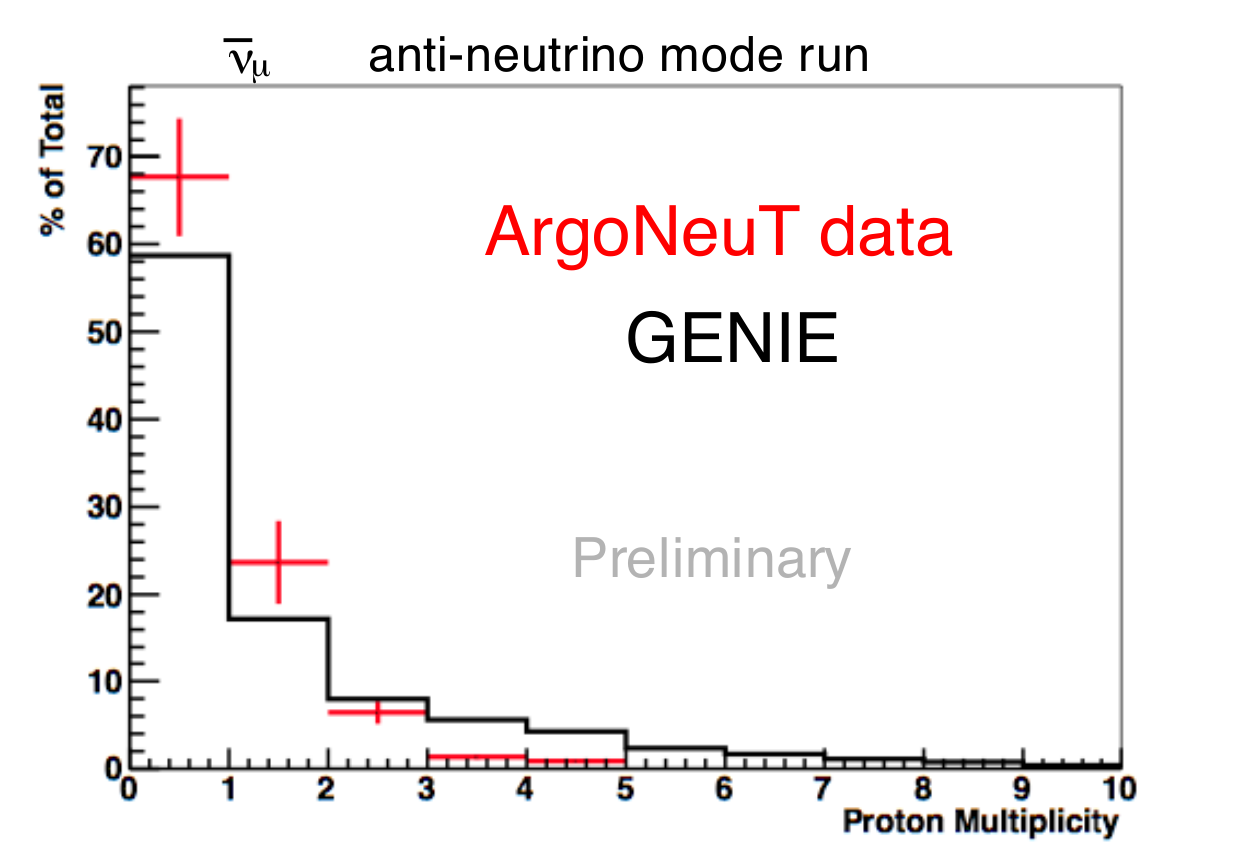}
\end{center}
\caption{\label{0pi} Distributions of proton multiplicity for $1\mu+Np+0\pi$ sample compared with {\sc genie} predictions. Right panel shows $\nu_{\mu}$ distributions and left panel shows $\bar{\nu}_{\mu}$ distributions. Both data and {\sc genie} predictions are normalized to unit area.}
\end{figure}

Figure~\ref{0pi} shows the distributions of proton multiplicity for $1\mu+Np+0\pi$ sample in comparison with {\sc genie} predictions. The {\sc genie} predictions have longer tails compared to ArgoNeuT data. This information is useful to improve the neutrino event generators.

\section{Conclusions}
ArgoNeuT was the first LArTPC placed in a low energy neutrino beam and it provides invaluable experience for future experiments that employ LArTPCs. We have measured the $\nu_{\mu}$ and $\bar{\nu}_{\mu}$ CC inclusive cross sections on argon. We also developed techniques to measure topological cross sections, which have yields results that are useful for understanding nuclear effects in neutrino-nucleus interactions. Currently there are many ongoing ArgoNeuT analyses. We are searching for events with one muon and two protons where the two protons go back-to-back. This would be an indication for nucleon-nucleon correlation in neutrino scattering. Other analyses include a search for neutral hyperon production, electron/photon identification, measurements of NC $\pi^{0}$ cross sections, nuclear deexcitation $\gamma$s, coherent pion production, and $\nu_{e}$ measurements.

\end{document}